\documentclass{ws-procs975x65}

\def\beq{\begin{equation}}
\def\eeq{\end{equation}}
\def\be{\begin{eqnarray}}
\def\ee{\end{eqnarray}}

\begin{document}

\title{Testing the Kerr Paradigm with the Black Hole Shadow}

\author{Cosimo Bambi}

\address{Center for Field Theory and Particle Physics and Department of Physics, Fudan University,\\
Shanghai 200433, China\\
Theoretical Astrophysics, Eberhard-Karls Universit\"at T\"ubingen,\\
T\"ubingen 72076, Germany\\
E-mail: bambi@fudan.edu.cn}

\begin{abstract}
Within 5-10 years, submillimeter VLBI facilities will be hopefully able to image the ``shadow'' of SgrA$^*$. When a black hole is surrounded by an optically thin emitting medium, the boundary of the shadow corresponds to the apparent photon capture sphere and only depends on the background metric. An accurate determination of the shape of the shadow of SgrA$^*$ could constrain possible deviations from the Kerr solution. In combination with other measurements, these observations could test the Kerr black hole paradigm.
\end{abstract}


\bodymatter


\section{Introduction}

Black hole candidates are astrophysical compact objects that can be naturally interpreted as the Kerr black holes of general relativity and they could be something else only in the presence of new physics. However, there is not yet direct evidence that the spacetime geometry around these objects is described by the Kerr solution and, at the same time, the predictions of general relativity have been tested only in weak gravitational fields. Deviations from standard predictions may be expected for a number of reasons, ranging from classical extensions of general relativity to macroscopic quantum gravity effects at the black hole horizon.

The radiation emitted in the vicinity of black hole candidates is affected by the strong gravitational field around these objects and the study of specific features can potentially test the Kerr black hole paradigm\cite{review}. This is not an easy job, because there is typically a degeneracy among the parameters of the system. In particular, there is usually a strong correlation between the estimate of the spin of the compact object and possible deviations from the Kerr metric, namely it is difficult (or even impossible) to distinguish a Kerr black hole from a non-Kerr object with a different spin parameter\cite{x1,x2,x3}. This problem can be solved if it is possible to combine several measurements of the same object and break the parameter degeneracy.

SgrA$^*$, the supermassive black hole candidate at the center of our Galaxy, is a special source and potentially a unique laboratory to test the Kerr nature of black hole candidates. While there are currently no good observational data to measure the metric around SgrA$^*$, the situation may change soon, as SgrA$^*$ could be explored with a number of unprecedented data that, when combined together, may represent the ideal case to test the Kerr black hole paradigm\cite{sgra}. Promising observations are:
\begin{enumerate}
\item The observation of radio pulsars\cite{pulsar} (or even normal stars\cite{sstar}) with orbital period of a few months.
\item The direct observation of blobs of plasma orbiting at a few gravitational radii from SgrA$^*$\cite{hotspot}
\item The measurement of the spectrum of the accretion structure of SgrA$^*$\cite{nan}.
\item The detection of the shadow of SgrA$^*$\cite{falcke,s1,s2,s3,s4}.
\end{enumerate}
In what follows, I will briefly discuss the detection of the shadow.

\section{Black hole shadow}

The direct image of a black hole surrounded by an optically thin emitting medium is characterized by a dark area, called ``shadow'' in the literature\cite{falcke}. While the intensity map of the image depends on the properties of the accretion flow and on the emission mechanisms, the exact shape of the shadow corresponds to the apparent photon capture sphere, which is only determined by the metric of the spacetime. An accurate measurement of the boundary of the shadow can constrain possible deviations from the Kerr solution.

The Cardoso-Pani-Rico (CPR) metric can be used as a parametrization to test the Kerr metric~\cite{cpr-m}. In Boyer-Lindquist coordinates, the line elements reads
\be\label{eq-m}
ds^2 &=& - \left(1 - \frac{2 M r}{\Sigma}\right)\left(1 + h^t\right) dt^2
+ \frac{\Sigma \left(1 + h^r\right)}{\Delta + h^r a^2 \sin^2\theta} dr^2
+ \Sigma d\theta^2 \nonumber\\
&& + \sin^2\theta \left\{\Sigma + a^2 \sin^2\theta \left[ 2 \sqrt{\left(1 + h^t\right)
\left(1 + h^r\right)} - \left(1 - \frac{2 M r}{\Sigma}\right)
\left(1 + h^t\right)\right]\right\} d\phi^2 \nonumber\\
&& - 2 a \sin^2\theta \left[\sqrt{\left(1 + h^t\right)\left(1 + h^r\right)} 
- \left(1 - \frac{2 M r}{\Sigma}\right)\left(1 + h^t\right)\right] dt d\phi  \, , 
\ee
where $a = J/M = a_* M$ is the specific BH spin with the dimension of $M$, $\Sigma = r^2 + a^2 \cos^2 \theta$, $\Delta = r^2 - 2 M r + a^2$, and
\be
\hspace{-0.4cm}
h^t = \sum_{k=0}^{+\infty} \left(\epsilon_{2k}^t 
+ \epsilon_{2k+1}^t \frac{M r}{\Sigma}\right)\left(\frac{M^2}{\Sigma}
\right)^k\, , \quad
h^r = \sum_{k=0}^{+\infty} \left(\epsilon_{2k}^r
+ \epsilon_{2k+1}^r \frac{M r}{\Sigma}\right)\left(\frac{M^2}{\Sigma}
\right)^k \, .
\ee
There are two infinite sets of deformation parameters, $\{\epsilon_k^t\}$ and $\{\epsilon_k^r\}$ (at increasingly high order). Since the lowest order deformation parameters are already strongly constrained to recover the Newtonian limit and meet the Solar System constraints\cite{cpr-m}, in what follows I will consider the deformation parameters $\epsilon_3^t$ and $\epsilon_3^r$.

\begin{figure}
\begin{center}
\includegraphics[type=pdf,ext=.pdf,read=.pdf,width=5.0cm]{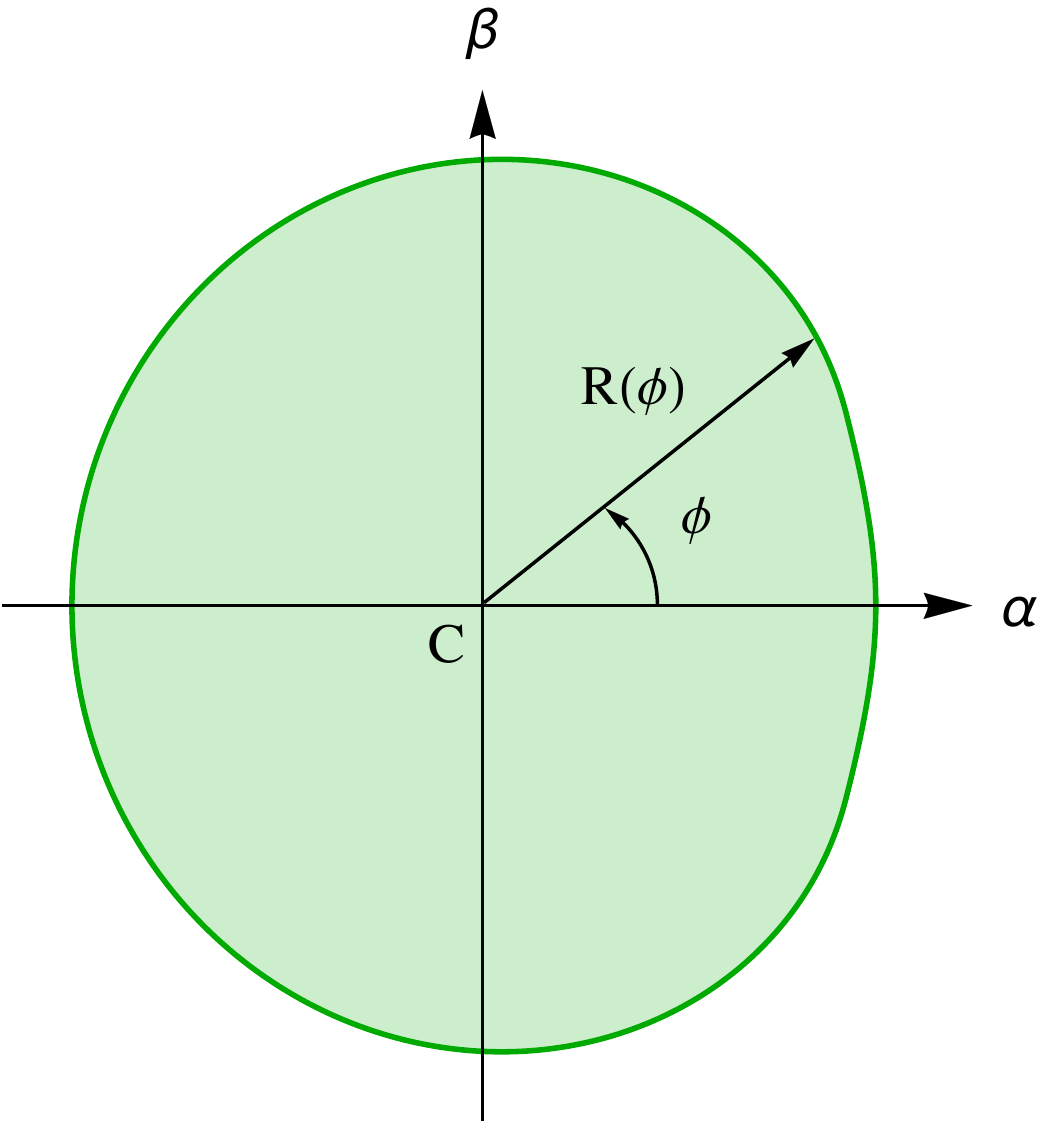}
\end{center}
\caption{$R(\phi)$ is the distance between the center of the shadow, C, and the point on the boundary of the shadow at the angle $\phi$. The function $R(\phi)/R(0)$ is used to describe the shape of the shadow. From Ref.~16. \label{fig1}}
\end{figure}

The shape of the shadow can be described following the approach of Ref.~\refcite{masoumeh}. First, we define the center C of the shadow as
\beq
X_{\rm C} = \frac{\int \rho(X,Y) X dX dY}{\int \rho(X,Y) dX dY} \, , \quad
Y_{\rm C} = \frac{\int \rho(X,Y) Y dX dY}{\int \rho(X,Y) dX dY} \, ,
\eeq
where $X$ and $Y$ are the Cartesian coordinates on the image plane of the observer, while $\rho (X,Y) = 1$ inside the shadow and $\rho (X,Y) = 0$ outside. We can then define the function $R(\phi)$ as the distance between C and the point on the boundary of the shadow at the angle $\phi$, as shown in Fig.~\ref{fig1} (see Ref.~\refcite{masoumeh} for more details). Since we do not have accurate measurements of the mass and of the distance of SgrA$^*$, we can only use the shape of the shadow to test the Kerr metric, not its size. In this case, we can use the function $R(\phi)/R(0)$, which completely describes the shape of the shadow.

\begin{figure}
\begin{center}
\includegraphics[type=pdf,ext=.pdf,read=.pdf,width=6.0cm]{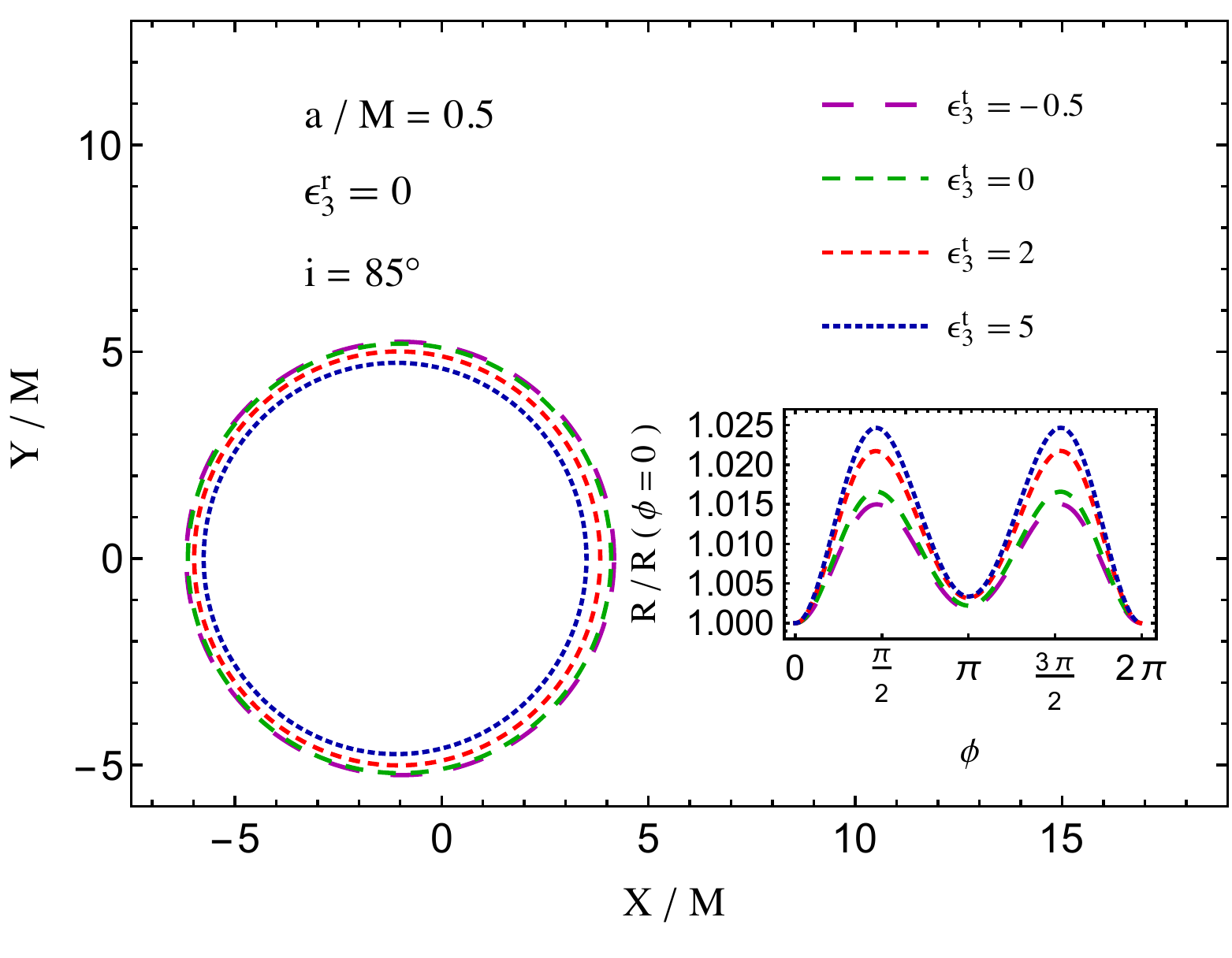}
\hspace{0.4cm}
\includegraphics[type=pdf,ext=.pdf,read=.pdf,width=6.0cm]{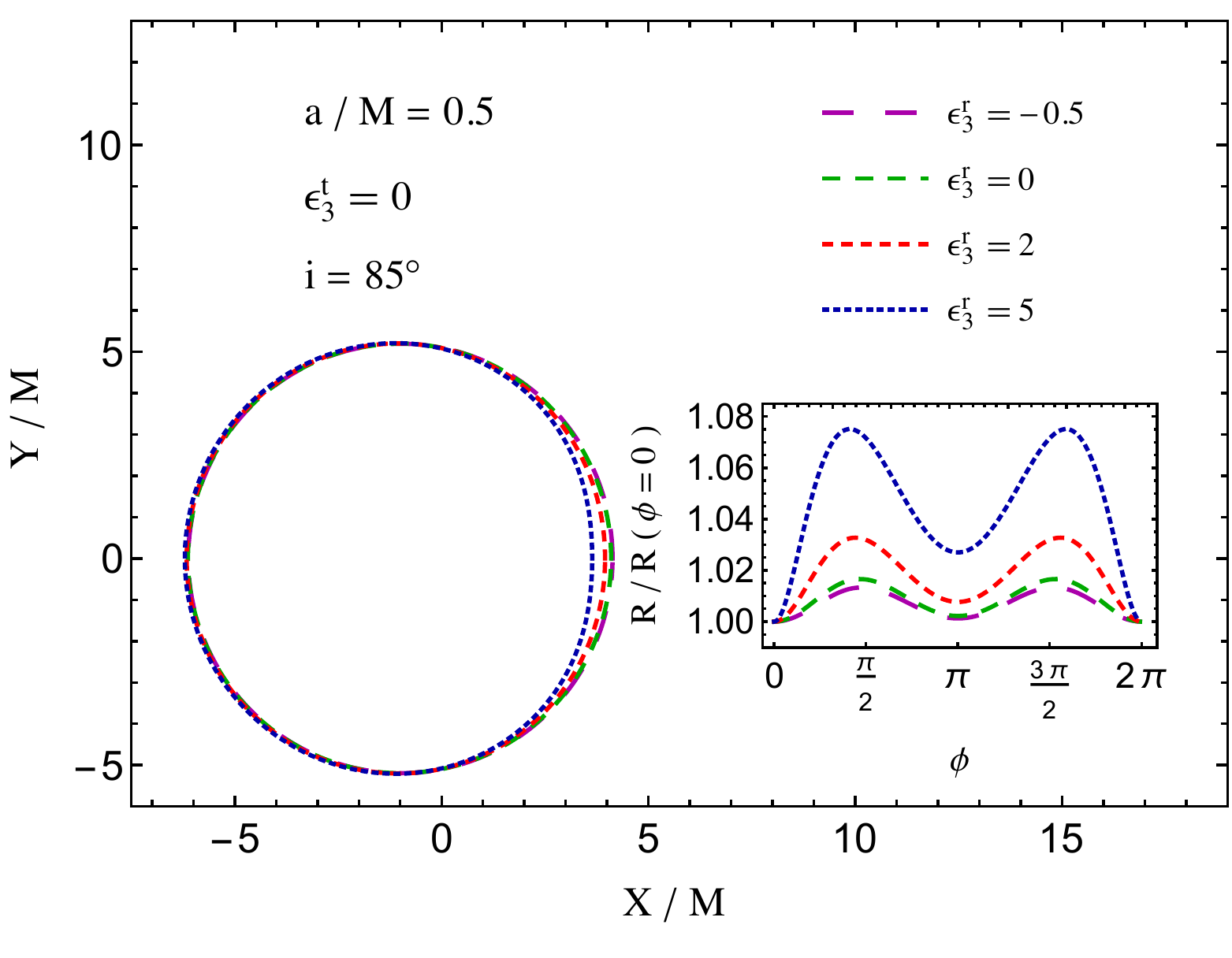} \\
\vspace{0.5cm}
\includegraphics[type=pdf,ext=.pdf,read=.pdf,width=6.0cm]{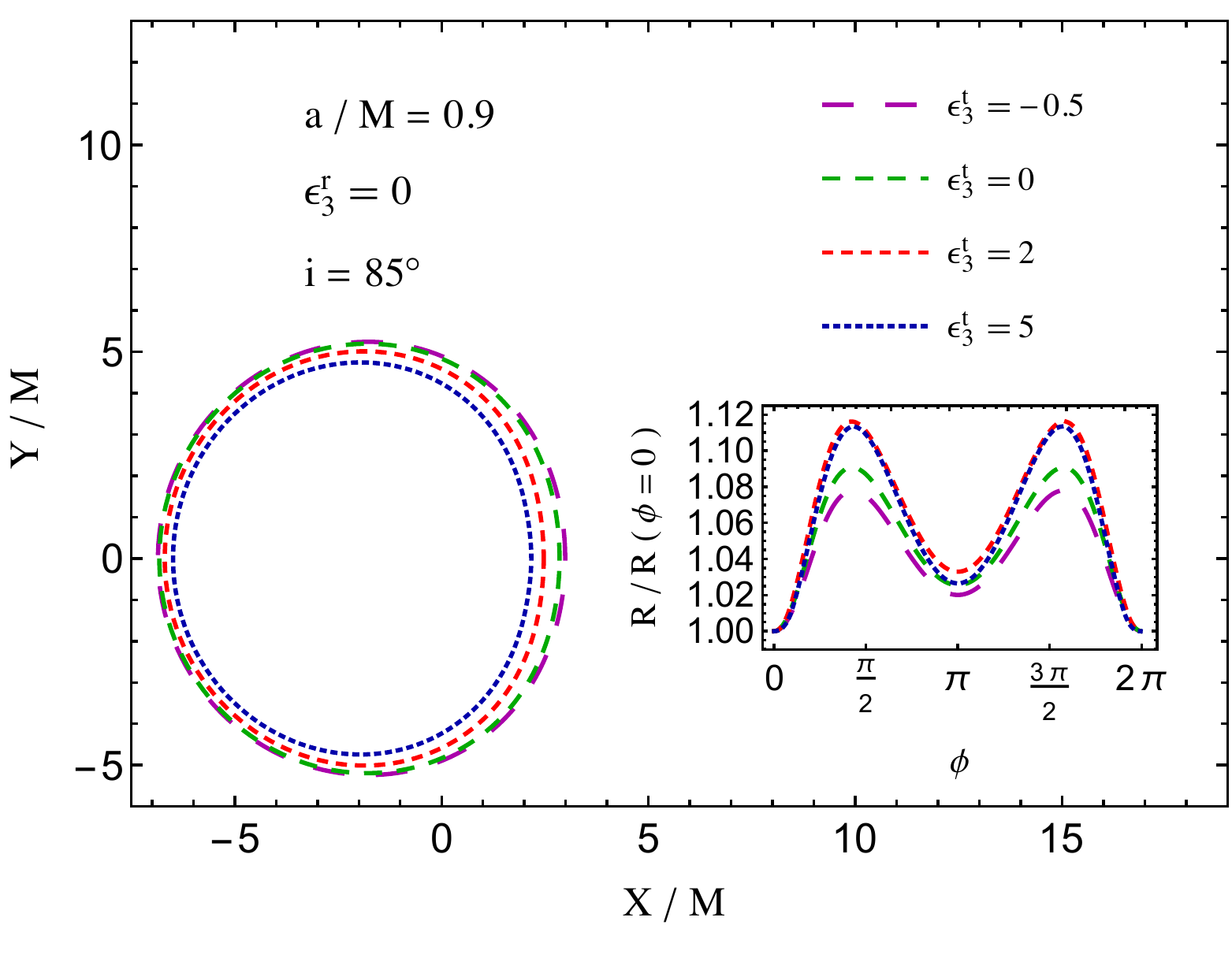}
\hspace{0.4cm}
\includegraphics[type=pdf,ext=.pdf,read=.pdf,width=6.0cm]{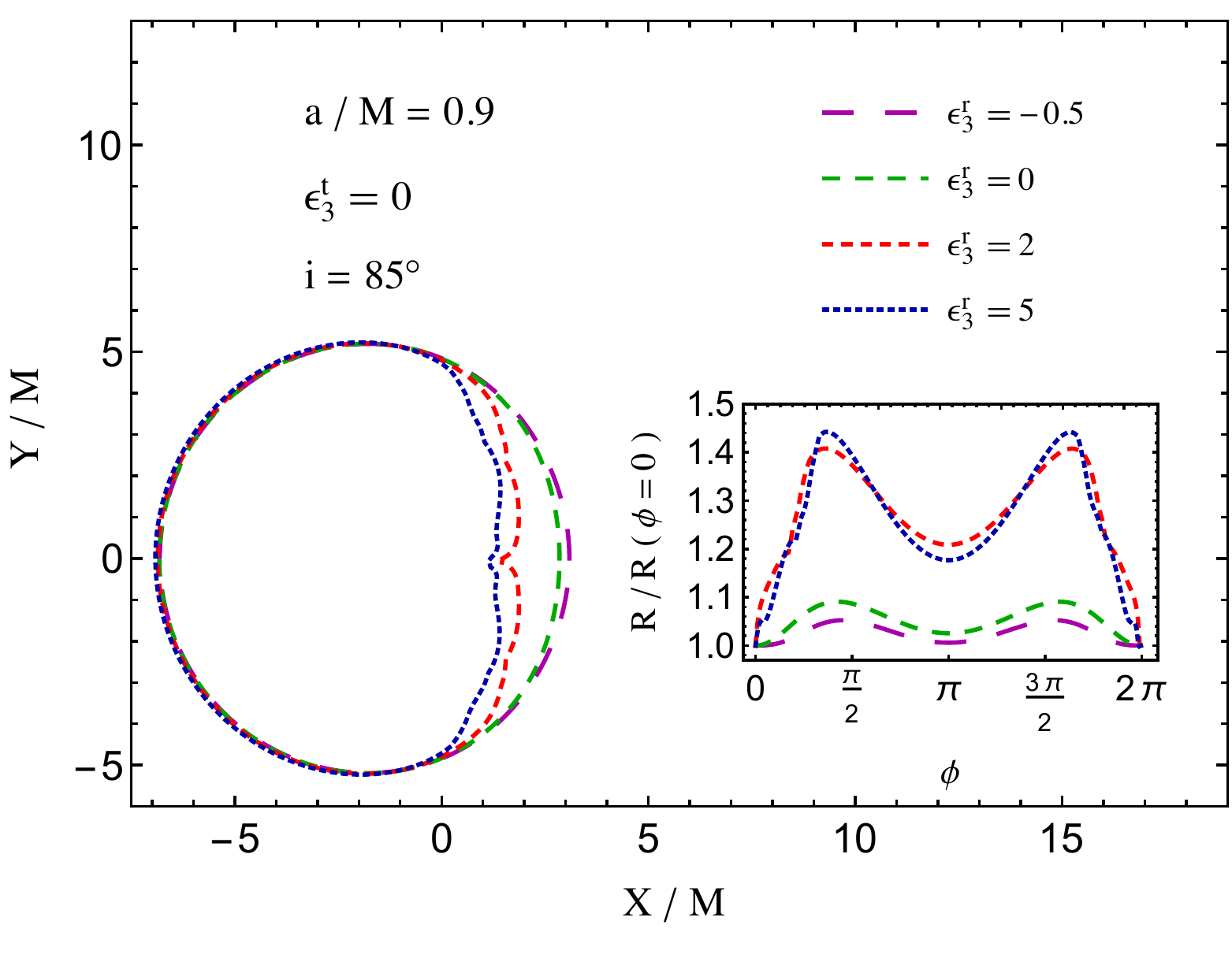}
\end{center}
\caption{Examples of shadows of CPR black holes with their function $R(\phi)/R(0)$ to describe the shape of the shadow. In the top panels, the spin parameter is $a_* = 0.5$, in the bottom panels it is $a_* = 0.9$. The left panels show the impact of $\epsilon^t_3$ assuming $\epsilon^r_3=0$. The right panels show the opposite case, with non-vanishing $\epsilon^r_3$ and $\epsilon^t_3=0$. The inclination angle is $i = 85^\circ$. From Ref.~16. See the text for more details. \label{fig2}}
\end{figure}

The shadows and the functions $R(\phi)/R(0)$ of some CPR black holes are shown in Fig.~\ref{fig2}. The spin parameter $a_*$ is 0.5 in the top panels and 0.9 in the bottom panels. In the left panels, $\epsilon^r_3 = 0$ and it is shown the impact on the shadow of $\epsilon^t_3$. In the right panel, we have the opposite case: $\epsilon^t_3 = 0$ and $\epsilon^r_3$ can vary. From Fig.~\ref{fig2}, we can see that $\epsilon^t_3$ mainly affects the size of the shadow, which increases (decreases) if $\epsilon^t_3$ decreases (increases). $\epsilon^r_3$ alters the shape of the shadow on the side of corotating orbits, while there are no appreciable effects in the other parts of the boundary of the shadow. The peculiar boundary appearing for $\epsilon^r_3 = 2$ and 5 in the bottom right panel is due to be non-trivial horizons of these black holes\cite{leonardo}

To be more quantitative and figure out if and how different shadows can be distinguished, we can proceed in the following way\cite{masoumeh}. We consider a ``reference model'', namely a black hole with a specific set of spin, deformation parameters, and viewing angle. For any black hole with parameters $(a_*, \epsilon^t_3, \epsilon^r_3, i)$, we can compute
\beq
S(a_*, \epsilon^t_3, \epsilon^r_3, i) = \sum_k \left( 
\frac{R(a_*, \epsilon^t_3, \epsilon^r_3, i; \phi_k)}{R(a_*, \epsilon^t_3, \epsilon^r_3, i; 0)} - 
\frac{R^{\rm ref} (\phi_k)}{R^{\rm ref} (0)} \right)^2 \, .
\eeq
where $R(a_*, \epsilon^t_3, \epsilon^r_3, i; \phi_k)$ is the function $R$ at $\phi = \phi_k$, $\{ \phi_k \}$ is a set of angles $\phi$ for which we consider a measurement, and $R^{\rm ref} (\phi_k)$ is the function $R$ of the reference model. The function $S$ can provide a simple estimate of the similarity between the shadow of the reference black hole and the shadow of the black hole with parameters $(a_*, \epsilon^t_3, \epsilon^r_3, i)$. It is related by the usual $\chi^2$ function by the approximate relation $\chi^2 \approx S/\sigma^2$, where $\sigma^2$ is the square of the error. If we can measure the boundary of the shadow with an uncertainty of 3\%, then $\sigma \approx 0.03$, and $\chi^2 \approx 1000 \; S$. In this case, the contour levels $\Delta \chi^2 = 3.53$, 8.03, and 14.16 corresponding, respectively, to the 1-, 2-, and 3-standard deviations for three degrees of freedom, become $S \approx 0.003$, 0.008, 0.014.

\begin{figure}
\begin{center}
\includegraphics[type=pdf,ext=.pdf,read=.pdf,width=6.0cm]{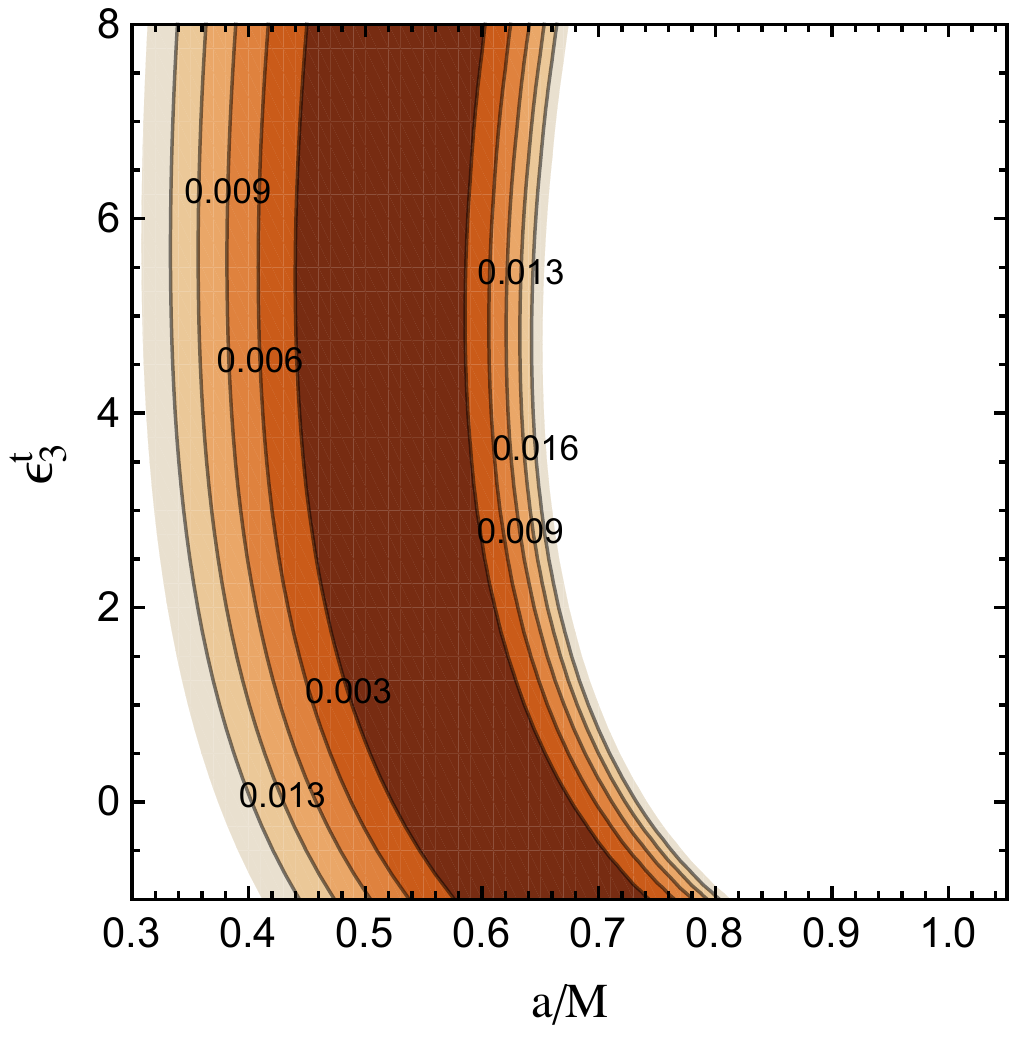}
\hspace{0.4cm}
\includegraphics[type=pdf,ext=.pdf,read=.pdf,width=6.0cm]{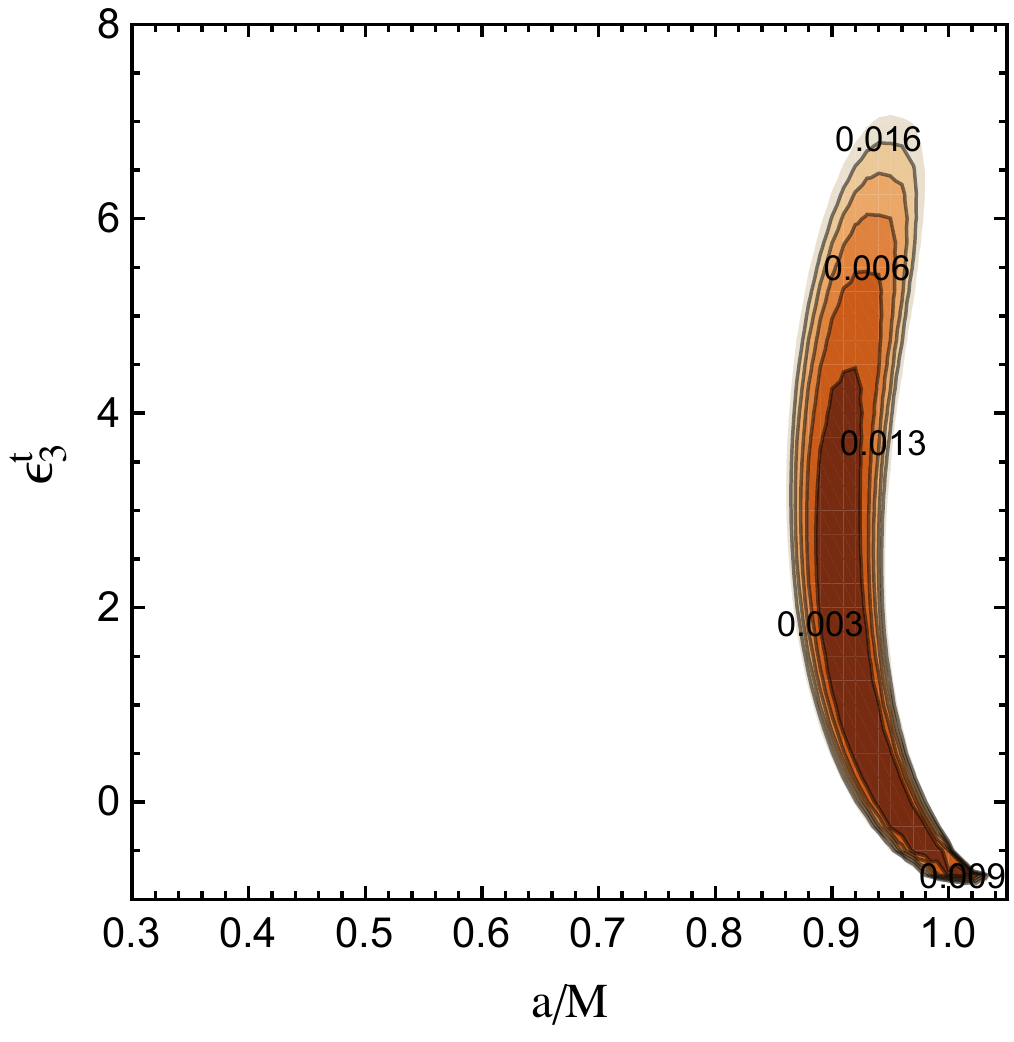} \\
\vspace{0.5cm}
\includegraphics[type=pdf,ext=.pdf,read=.pdf,width=6.0cm]{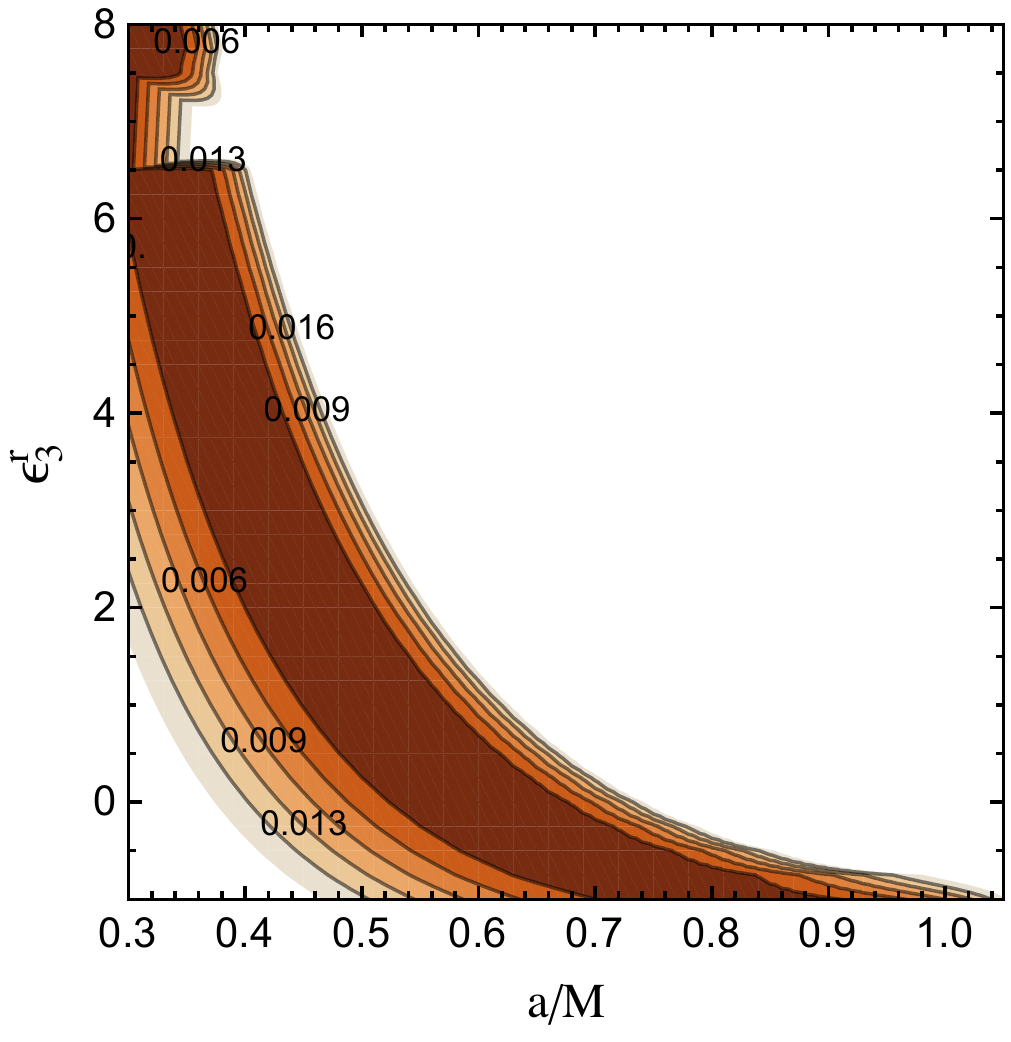}
\hspace{0.4cm}
\includegraphics[type=pdf,ext=.pdf,read=.pdf,width=6.0cm]{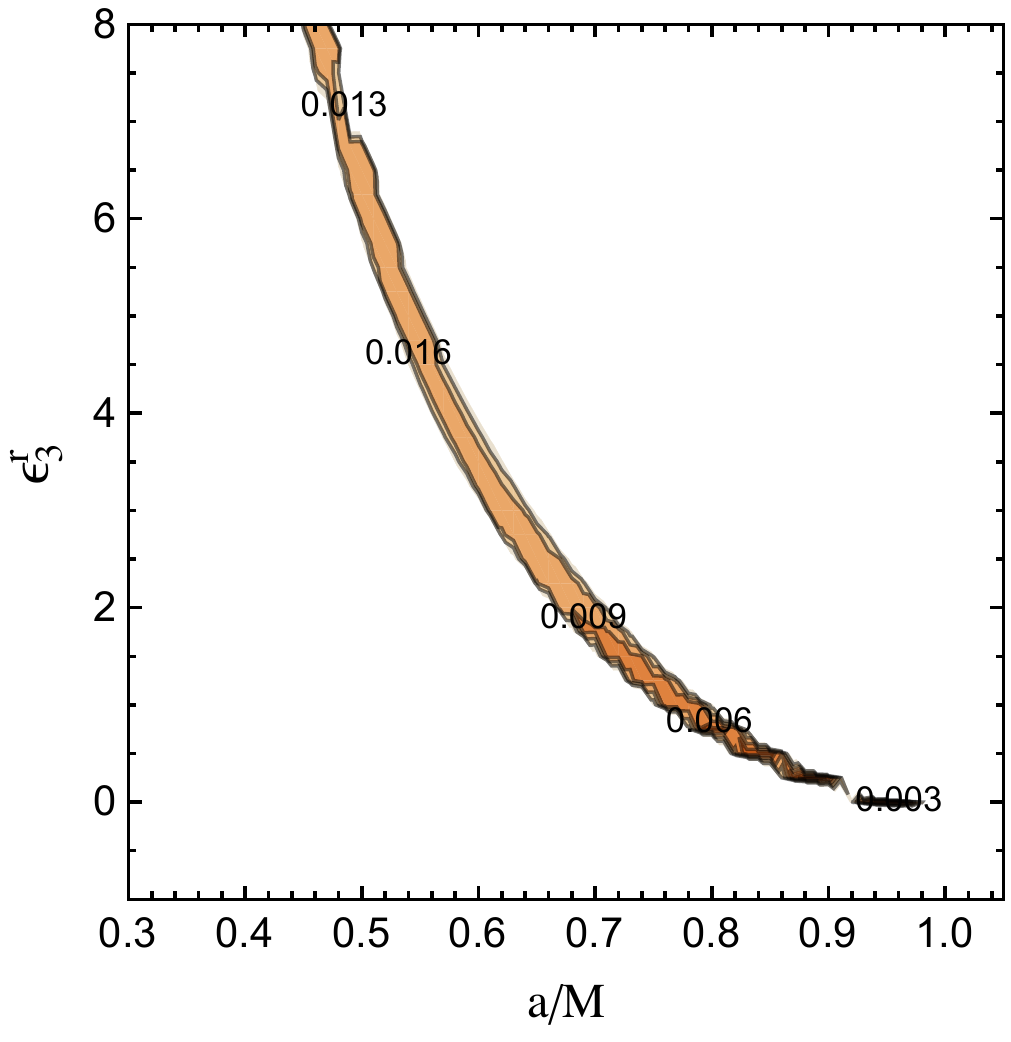}
\end{center}
\caption{Contour maps of $S$. In the left panels, the reference model is a Kerr black hole with spin parameter $a_* = 0.6$. In the right panels, the reference model is a Kerr black hole with spin parameter $a_* = 0.95$. The inclination angle of both reference models is $i = 80^\circ$. From Ref.~16. \label{fig3}}
\end{figure}

Fig.~\ref{fig3} shows some examples of contour maps of $S$. In the top panels, the only non-vanishing deformation parameter is $\epsilon^t_3$ and $\epsilon^r_3 = 0$. In the bottom panels, $\epsilon^t_3 = 0$ and $\epsilon^r_3$ may not vanish. The left panels are for a reference black hole with $a_* = 0.6$, $\epsilon^t_3 = \epsilon^r_3 = 0$, and $i = 80^\circ$. In the right panels, the reference black hole has $a_* = 0.95$, $\epsilon^t_3 = \epsilon^r_3 = 0$, and $i = 80^\circ$. In this plots, $S$ is minimized with respect to $i$, since it is not assumed that $i$ cannot be obtained from independent measurements. See Ref.~\refcite{masoumeh} for more details.

\section{Concluding remarks}

Tests of the Kerr metric are usually affected by a strong parameter degeneracy, in particular it is usually problematic to measure the spin of the compact object and, at the same time, constrain possible deviations from the Kerr solution. The combination of different measurements of the same source is the natural way to break the parameter degeneracy. From this point of view, SgrA$^*$ is a promising source to test the Kerr black hole paradigm. The combination of precise astrometric measurements in the weak field (pulsars, stars), astrometric observations of blobs of plasma in the strong field, accurate measurements of the spectrum of its accretion structure, and the detection of its shadow may provide a quite unique opportunity to test the Kerr metric.

The shadow is one of the possible measurements. While there is a common consensus on the detectability of the shadow of SgrA$^*$ in the next 5-10 years\cite{doeleman}, it is not yet clear the level of precision that can be reached. An accurate measurement of the boundary of the shadow would correspond to the detection of the apparent photon capture sphere of SgrA$^*$ and could constrain possible deviations from the Kerr background as shown in Fig.~\ref{fig3}. It is unlikely that these data will be able to test the Kerr metric, at least if they do not show some peculiar feature that is not expected in the Kerr case, but they could be useful in combination with other measurements.


\section*{Acknowledgments}

This work was supported by the NSFC grant No.~11305038, the Shanghai Municipal Education Commission grant No.~14ZZ001, the Thousand Young Talents Program, Fudan University, and the Alexander von Humboldt Foundation.


\end{document}